\documentclass[aps,prl,preprint,floatfix,showpacs,groupedaddress,superscriptaddress,longbibliography]{revtex4-1}
\usepackage{graphicx}
\usepackage{color}
\usepackage{epstopdf}

\begin{document}
\title{Thermal conductivity of the Kondo semiconductor CeRu$_4$Sn$_6$}

\author{J H\"anel}
\email{haenel@ifp.tuwien.ac.at}
\affiliation{Institute of Solid State Physics, Technische Universit\"at Wien, Wiedner Hauptsr. 8-10, 1040 Wien, Austria}
\author{M Taupin}
\affiliation{Institute of Solid State Physics, Technische Universit\"at Wien, Wiedner Hauptsr. 8-10, 1040 Wien, Austria}
\author{M Ikeda}
\affiliation{Institute of Solid State Physics, Technische Universit\"at Wien, Wiedner Hauptsr. 8-10, 1040 Wien, Austria}
\author{V Martelli}
\affiliation{Institute of Solid State Physics, Technische Universit\"at Wien, Wiedner Hauptsr. 8-10, 1040 Wien, Austria}
\affiliation{Centro Brasileiro de Pesquisas F\'{i}sicas, Rua Doutor Xavier Sigaud 150, CEP 22290-180, Brazil}
\author{P Tome\v{s}}
\affiliation{Institute of Solid State Physics, Technische Universit\"at Wien, Wiedner Hauptsr. 8-10, 1040 Wien, Austria}
\author{A Prokofiev}
\affiliation{Institute of Solid State Physics, Technische Universit\"at Wien, Wiedner Hauptsr. 8-10, 1040 Wien, Austria}
\author{S Paschen}
\affiliation{Institute of Solid State Physics, Technische Universit\"at Wien, Wiedner Hauptsr. 8-10, 1040 Wien, Austria}

\begin{abstract}
We report measurements of the thermal conductivity $\kappa$ on single crystalline CeRu$_4$Sn$_6$ in the temperature range between 80\,mK and 80\,K, along the main crystallographic directions.
$\kappa$ is phonon-dominated in the whole temperature range and is found to be essentially isotropic. At low temperatures the data are well approximated by $\kappa \propto T^2$, which is attributed to a predominant scattering of phonons on electrons.
We describe the data with a Callaway fit in the whole temperature range giving good agreement at low and high temperatures.

\end{abstract}

\maketitle
\section{Introduction}
Kondo insulators or Kondo semiconductors (KSs) are strongly correlated electron systems where a narrow (pseudo)gap opens at low temperatures in the electronic density of states (DOS) at the Fermi level~\cite{Aeppli92}.
This is a result of the hybridization of localized $f$ electrons with the conduction $p$ or $d$ electrons~\cite{Riseborough00}.
Even though known since decades~\cite{Aeppli92}, KSs remain at the forefront of condensed matter research.
Recently it has been suggested that KSs may be candidates for strongly correlated materials with topologically protected surface states~\cite{Dzero10,Dzero12,Alexandrov13}.
So far, experiments along these lines have focused on the cubic KS SmB$_6$~\cite{Cooley99,Kim14}. \\
Here CeRu$_4$Sn$_6$ comes into play.
It has a tetragonal, non-centrosymmetric crystal structure ($I\bar{4}2m$)~\cite{Poettgen97} and therefore can shed light on the role crystal symmetry plays.
It was first synthesized in 1992~\cite{Das92} and a development of a narrow gap was found in several physical properties~\cite{Strydom05,Bruening10,Paschen10}.
Measurements on single crystals revealed large anisotropy in resistivity, specific heat, and thermopower~\cite{Paschen10,Winkler12,Haenel14}.
Optical conductivity measurements and LDA+DMFT calculations attributed this to a gapping in the tetragonal plane while the out-of-plane direction stays weakly metallic~\cite{Guritanu13}. \\
Here we report on the thermal conductivity $\kappa$ of CeRu$_4$Sn$_6$ single crystals, both along and perpendicular to the out-of-plane direction $c$, in the temperature range of 80\,mK to 80\,K.

\section{Experimental}
\label{exp}
The polycrystalline starting material for the crystal growth was synthesized by melting first Ce and Ru, and subsequently the resulting Ce/Ru alloy with Sn in a horizontal water cooled copper boat using high-frequency heating. 
The purity of the starting materials was 99.99\% for Ce (Ames Lab) and Ru, and 99.9999\% for Sn. 
The single crystal growth was performed by the floating zone melting technique using optical heating in a four-mirror furnace of Crystal Corporation.
All steps were performed under Ar 99.9999\% protective atmosphere after purging several times with Ar. 
Since CeRu$_4$Sn$_6$ melts incongruently we used a self-flux technique \cite{Paschen10,Prokofiev12}. \\
The low-temperature (80\,mK~-~2\,K) thermal conductivity was measured in a $^3$He/$^4$He dilution refrigerator with a home-built setup employing a two-thermometer-one-heater technique using RuO$_2$ thick-film resistors as heater and thermometers.
The measurements in the range 2\,K to 80\,K were carried out in a $^4$He bath cryostat with thermocouples to measure the temperature gradient.

\section{Physical properties}
\label{phys}
\begin{figure}[h]
\centering
\includegraphics[width=28pc]{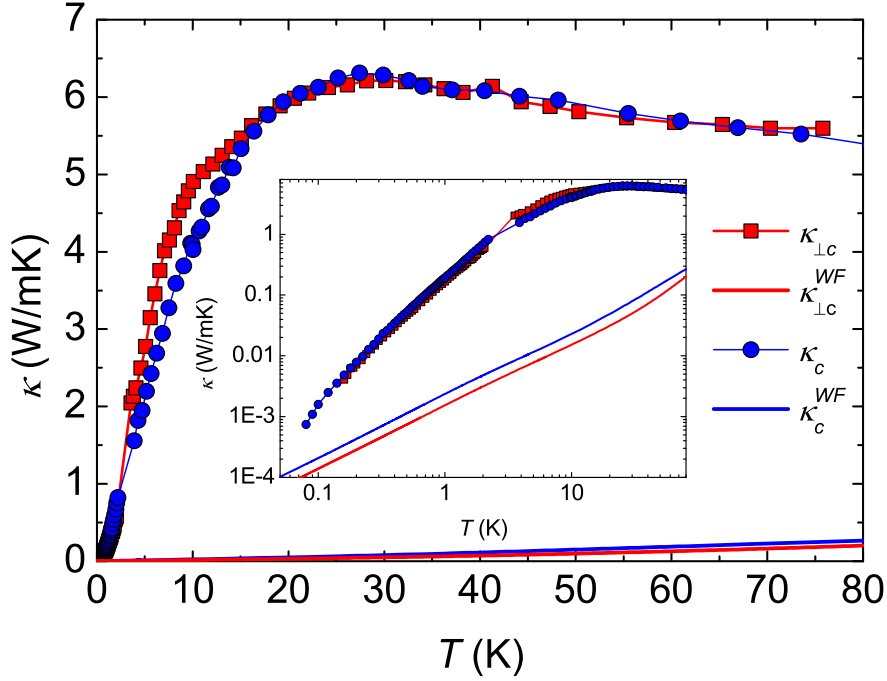}
\caption{\label{fig:k}Thermal conductivity $\kappa$ of CeRu$_4$Sn$_6$ as a function of temperature $T$ measured parallel and perpendicular to the out-of-plane direction $c$. The electronic part of the thermal conductivity $\kappa^{\rm{WF}}$, as obtained from resistivity $\rho$~\cite{Winkler12} via the Wiedemann-Franz law, is plotted as solid lines. The inset shows the same data on a double-logarithmic scale.}
\end{figure}

Figure\,\ref{fig:k} shows the thermal conductivity $\kappa$ as a function of temperature $T$ in the whole temperature range investigated, on both a linear (main panel) and a double-logarithmic (inset) scale.
$\kappa$ was measured along the out-of-plane direction $c$ (called $\kappa_c$) and within the tetragonal plane, perpendicular to $c$ ($\kappa_{\perp c}$).
We calculated the electronic part of the thermal conductivity $\kappa^{\rm{WF}}$ from the electrical resistivity $\rho$\,\cite{Winkler12} employing the Wiedemann-Franz law $\kappa^{\rm{WF}} = L_0 T / \rho $ with the Lorenz number $L_0=2.44 \times 10^{-8}$\,W$\Omega$K$^{-2}$.
$\kappa$ is essentially isotropic, in marked contrast to other physical properties such as electrical resistivity~\cite{Winkler12}, thermopower~\cite{Haenel14}, and optical conductivity~\cite{Guritanu13}. \\
At low temperatures $\kappa$ increases strongly with increasing temperature towards a maximum of roughly $\kappa_{\rm{max}} \approx$ 6.3\,Wm$^{-1}$K$^{-1}$ at $T_{\rm{max}} = $ 28\,K.

\begin{figure}[h]
\centering
\includegraphics[width=28pc]{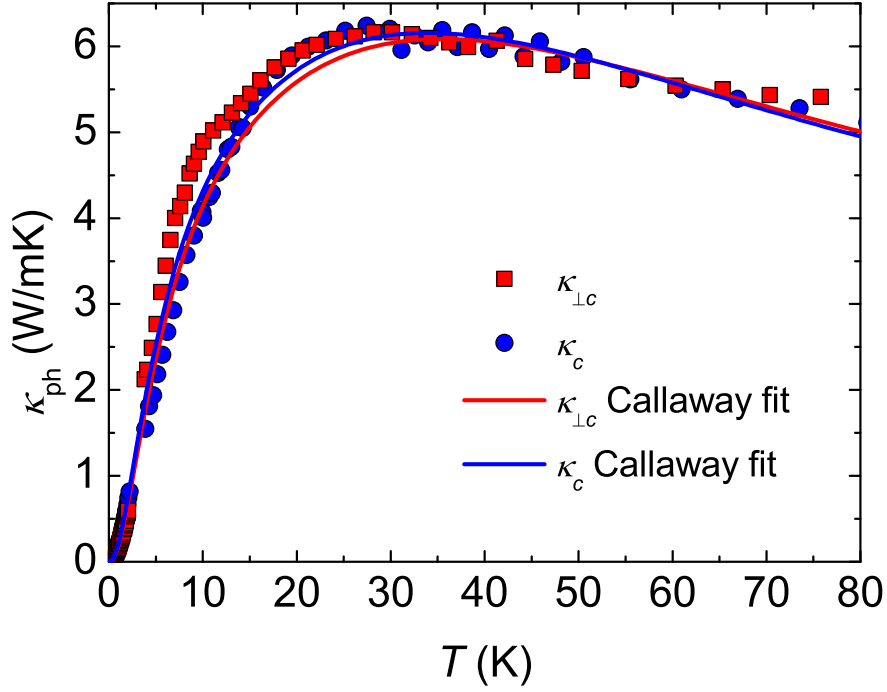}
\caption{\label{fig:kCallaway}Phononic part of the thermal conductivity $\kappa_{\rm{ph}}$ of CeRu$_4$Sn$_6$ along and perpendicular to the $c$ direction as a function of temperature $T$. Solid lines are Callaway fits to the data (see text).}
\end{figure}
$\kappa^{\rm{WF}}$ is much lower than the total $\kappa$ and thus $\kappa$ is clearly phonon-dominated as expected in KSs. 
We estimate the phononic contribution $\kappa_{\rm{ph}}$ as $\kappa_{\rm{ph}} = \kappa - \kappa^{\rm{WF}}$.
The result is plotted in Fig.\,\ref{fig:kCallaway}, together with least-square fits to the Callaway model~\cite{Callawy59}
\begin{equation}
\kappa_{ph} = \frac{k_B^4 \cdot T^3}{2 \cdot \pi^2 \cdot v_g \cdot \hbar^3} \cdot \left[\int^{\frac{\theta_D}{T}}_0 {\tau \cdot \frac{x^4 \cdot e^x}{\left(e^x-1\right)^2} \cdot dx} \right] \quad ,   
\label{eq:callaway}  
\end{equation}
where $v_g$ is the sound velocity, $\theta_D$ is the Debye temperature, $\tau$ is the phonon relaxation rate, $x = \frac{\hbar \omega}{k_B T}$, and the other symbols have the usual meaning.
Different scattering mechanisms influence $\tau$ according to the Matthiesen's rule
\begin{equation}
\frac{1}{\tau} = \sum_i{\frac{1}{\tau_i}} \quad.
\label{eq:matt}
\end{equation}
The various scattering mechanisms have different dependencies on the frequency $\omega$ and subsequently lead to different temperature dependencies of $\kappa_{\rm{ph}}$.
We have taken phonon-electron, -boundary, -defect and Umklapp scattering processes into account.
Note that we have neglected N processes here which would lead to a correction term in Eqn.\,\ref{eq:callaway}.
However the influence of these processes is expected to be small in materials with complex crystal structure.
Table\,\ref{tab:fits} lists the results of the fits. \\
\begin{center}
\begin{table}[h]
\caption{\label{tab:fits}Fit parameters of the Callaway fits, where the scattering rates are phonon-electron scattering $\tau^{-1}_{\rm{pe}} = A \cdot \omega^2$, phonon-boundary scattering $\tau^{-1}_{\rm{b}} = B$, phonon-defect scattering $\tau^{-1}_{\rm{pd}} = C \cdot \omega^4 $, and Umklapp scattering $\tau^{-1}_{\rm{U}} = D \cdot \omega^2 \cdot T \cdot exp(\frac{\theta_D}{-3T})$. The values in the table are the constants $A$, $B$, $C$, and $D$.}
\centering
\begin{tabular}{lcccc}
\hline
Direction & $A$ & $B$ & $C$ & $D$\\
  & (s) & (s$^{-1}$) & (s$^3$) & (sK$^{-1}$)\\
\hline
$c$ & \quad $2.2 \times 10^7$ \quad & \quad $2.8 \times 10^6$ \quad  & \quad 7250 \quad & \quad $9.16 \times 10^4$ \\
$\perp c$ & \quad $2.5 \times 10^7$ \quad   & \quad $3.0 \times 10^6$ \quad  & \quad 6878  \quad  & \quad $9.06 \times 10^4$ \\
\hline
\end{tabular}
\end{table}
\end{center}
\begin{figure}[h]
\centering
\includegraphics[width=28pc]{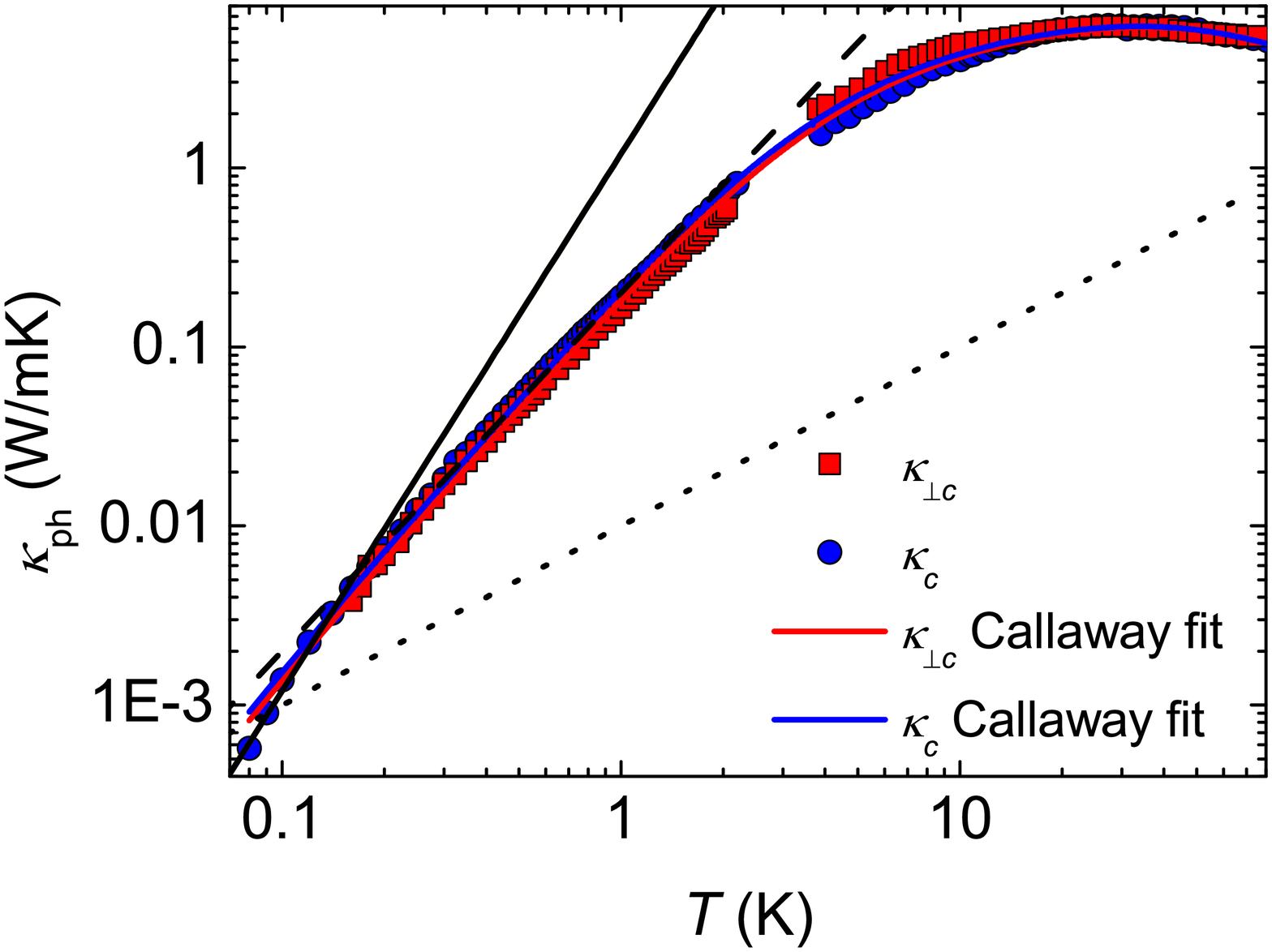}
\caption{\label{fig:kCallawayLog}Phonon thermal conductivity $\kappa_{\rm{ph}}$ of CeRu$_4$Sn$_6$ along and perpendicular to the $c$ direction as a function of temperature $T$ on a double-logarithmic scale. Solid red and blue lines are fits to the data according to the Callaway model~\cite{Callawy59}. Black lines indicate different power laws: $T^3$ (solid), $T^2$ (dashed), and linear-in-$T$ (dotted).}
\end{figure}
Figure\,\ref{fig:kCallawayLog} shows the same data as Fig.\,\ref{fig:kCallaway} on a double-logarithmic scale, which reveals the different scattering contributions most clearly.
Straight lines indicate different power laws.
The solid line shows $T^3$ behaviour typical for phonon-boundary scattering.
The dashed line indicates $T^2$ behaviour which is seen for phonon-electron scattering, and the dotted line is linear in $T$. \\
The data follow the dashed line very closely in the range of 200\,mK to 2\,K.
Thus the dominant scattering mechanism in this range is identified as phonon-electron scattering.
At lower temperatures $\kappa_{\rm{ph}}$ shows a $T^3$ behaviour. 
The fit parameters of the Callaway fits give the boundary scattering rates $\tau_{\rm{b},\perp c}^{-1} = 2.8 \times 10^6\,\rm{s}^{-1}$ and $\tau_{\rm{b},c}^{-1} = 3.0 \times 10^6\,\rm{s}^{-1}$ for the measurement perpendicular and parallel to $c$, respectively.
With a Debye temperature $\theta_D=250$\,K~\cite{Bruening10} the sound velocity can be computed to be $v_g = 2316$\,m/s.
With this $v_g$ the scattering times correspond to the mean free paths $l_{\perp c} = 0.83$\,mm and $l_{c} = 0.77$\,mm.
This is close to the smallest dimension of the samples and we therefore conclude that the phonons are predominantly scattered off the sample boundaries in this temperature range. \\
The onset of phonon-electron scattering, with the scattering rate $\tau_{\rm{pe}}^{-1}$, at higher temperatures leads to a further reduction of the thermal conductivity. According to Pippard's theory~\cite{Pippard55}, $\tau_{\rm{pe}}^{-1}$ is proportional to the charge carrier concentration $n$. 
From Hall effect measurements $n$ was determined to be isotropic at both high and low temperatures~\cite{WinklerTh13} in agreement with the isotropic behaviour of $\kappa$ in these temperature ranges. However, at intermediate temperatures, signatures of the Kondo gap opening are anisotropic ~\cite{Guritanu13}. This may explain the small anisotropy seen in the thermal conductivity around 10\,K (hump in $\kappa_{\perp c}$, see Fig.\,\ref{fig:kCallaway}). As the gap opening is more pronounced for the plane perpendicular to $c$, the effect of electrons freezing out and thus disappearing as scattering partners for the phonons is more pronounced for $\kappa_{\perp c}$, which leads to slightly enhanced $\kappa_{\perp c}$ values. Similar anisotropies have been found in the KSs CeNiSn and CeRhSb and have likewise been attributed to anisotropies in the energy gap~\cite{Paschen00,Sera97,Isikawa91}.

Above 2 K, $\kappa_{\rm{ph}}$ deviates from the $T^2$ law to lower values. Here scattering on point defects and U processes become the dominant effects. The values of the prefactors for these processes, as listed in Tab.\,\ref{tab:fits}, are similar to what was found in other systems~\cite{Yan13}. However, for a quantitative discussion it is necessary to know the phonon band structure which is not available in this system. In addition, other aspects that play an important role in CeRu$_4$Sn$_6$ and are neglected in a simple Callaway model may come into play at elevated temperatures.

\section{Conclusion}
\label{con}
In summary we have investigated single-crystalline CeRu$_4$Sn$_6$ by means of thermal conductivity measurements in the temperature range of 80\,mK to 80\,K. 
The data are essentially isotropic in contrast to findings for other physical properties. The reason is that the thermal conductivity is dominated by its phonon contribution, even at the lowest temperature where the behaviour
can be well explained by boundary and phonon-electron scattering.
This gives further support to the previous conjecture that the strongly anisotropic electrical resistivity is mostly due to anisotropies in the charge carrier mobility, with the charge carrier concentration being essentially isotropic.

\section{Acknowledgement}
We gratefully acknowledge financial support from the Austrian Science Fund (FWF projects I 623-N16, W1243, and I 2535-N27) and from the European Research Council (ERC grant 227378).
VM acknowledges the FAPERJ (Nota 10).

\bibliography{Haenel_SCES16_b.bbl}

\end{document}